\documentstyle[psfig]{mn}
\def\snra{G296.8--00.3}
\newcommand\HI{H\,{\sc i}}
\newcommand\HII{H\,{\sc ii}}
\newcommand\kms{km~s$^{-1}$}

\begin{document}
\label{firstpage}
\title[Radio observations of SNR~\snra]
{Radio continuum and \HI\ observations of 
supernova remnnant \snra}
\author[B. Gaensler, R. Manchester \& A. Green]
{B. M. Gaensler$^{1,2}$, R. N. Manchester$^2$ and A. J. Green$^1$ \\
$^1$Astrophysics Department, School of
Physics A29, University of Sydney, NSW 2006, Australia; \\
b.gaensler@physics.usyd.edu.au, a.green@physics.usyd.edu.au \\ 
$^2$Australia Telescope National Facility, CSIRO, PO Box 76,
Epping, NSW 2121, Australia; rmanches@atnf.csiro.au}

\pagerange{\pageref{firstpage}--\pageref{lastpage}}
\pubyear{1997}

\maketitle
\begin{abstract}

We present Australia Telescope Compact Array (ATCA) observations of the
supernova remnant (SNR) \snra. A 1.3-GHz continuum image shows
the remnant to have a complex multi-shelled appearance, with an unusual
rectangular strip running through its centre.  \HI\ absorption yields a
kinematic distance to the remnant of $9.6\pm0.6$~kpc, from which we
estimate an age in the range $(2-10) \times 10^3$~yr. The ATCA's continuum
mode allows a measurement of the Faraday rotation across the band, from
which we derive a mean rotation measure towards the SNR of
430~rad~m$^{-2}$.
We consider possible explanations for the morphology of \snra, and
conclude that it either has a biannular structure, as might be
produced through interaction with an asymmetric progenitor wind,
or that its appearance is caused by the effects of the surrounding
interstellar medium.

We argue that the adjacent pulsar J1157--6224 is at a similar distance
as the SNR, but that a physical association is highly unlikely.  The
pulsar is the only detectable source in the field in circular
polarization, suggesting a method for finding pulsars during aperture
synthesis.

\end{abstract}

\begin{keywords}
circumstellar matter --- 
pulsars: individual (J1157--6224) --- 
radio lines: ISM ---
shock waves --- 
supernova remnants:  individual (\snra) --- 
techniques: polarimetric
\end{keywords}

\section{Introduction}
\label{sec_intro}

The expanding shock front produced by a supernova (SN) explosion is
best delineated in the radio continuum, where we observe the
synchrotron emission from ultra-relativistic electrons at or near the
expanding shock. Few supernova remnants (SNRs) appear circular and
undistorted, but whether this distortion is a result of the distribution of the
ejecta driving the shock front, the mass-loss history of the
progenitor, the structure of the surrounding interstellar medium (ISM)
or some complex combination of
all of these, is a difficult question to answer. Of the 215 known SNRs
in the Galaxy \cite{gre96b}, many have now been observed at sub-arcmin
resolution at radio frequencies (e.g.\ Whiteoak \& Green
1996\nocite{wg96}). However, without knowledge of their physical
properties interpretation of their appearance is difficult.

Although detected in early radio surveys of the Galactic Plane
\cite{hil68,td69}, \snra\ was first identified as a supernova remnant
by Large \& Vaughan \shortcite{lv72} on the basis of its spectral index
and rough shell-like morphology. These authors claimed an association with
the pulsar J1157--6224 (B1154--62), 13 arcmin from the remnant's
centre, which has since been considered unlikely on the basis of the
pulsar's large timing age of $1.6 \times 10^6$~yr \cite{gj95c,jkww96}.
A recent higher resolution image (43 $\times$ 49 arcsec) of this
SNR has been made at 843~MHz by Whiteoak \& Green \shortcite{wg96},
showing a complicated multi-ringed structure, brightest to the
north-west.

In X-rays, Hwang \& Markert \shortcite{hm94} report a 4-$\sigma$
detection of SNR~\snra\ with {\em ROSAT}\ PSPC.  They ascribe the
weakness of the X-ray emission to significant absorption along the line
of sight, and consequently estimate a distance of $\sim10$~kpc.  There
is no detectable infrared emission associated with the SNR
\cite{are89,sfs92}, and examination of the Digitized Sky Survey and of the
corresponding ESO~R field shows no optical counterpart.

As part of a programme to determine the basic properties of southern
SNRs with unusual features, we present in this paper 1.3-GHz continuum
and \HI\ absorption observations towards SNR~\snra.  We describe our
observations and analysis in Section~\ref{sec_obs} and present our
results in Section~\ref{sec_results}.  In Section~\ref{sec_discuss} we
derive physical properties for the SNR and discuss possible causes for
the remnant's unusual morphology. We also consider the possibility of
an association between SNR~\snra\ and PSR~J1157--6224.

\section{Observations and data reduction}
\label{sec_obs}

Observations were made with the Australia Telescope Compact Array
(ATCA; Frater, Brooks \& Whiteoak 1992\nocite{fbw92}), a synthesis 
telescope near Narrabri, New South Wales. The ATCA
consists of five 22-m-diameter antennae located on
a 3-km east-west track, with a fixed sixth antenna located 3~km further west.
A pointing
centre RA (J2000) $11^{\rm h}58^{\rm m}43^{\rm s}$, Dec (J2000)
$-62\degr34\arcmin29\arcsec$ was observed with three different array
configurations, as shown in Table~\ref{tab_observations}.  For
the compact configurations used here, correlations involving 
the sixth antenna were not used. Continuum
and \HI\ line observations were made simultaneously; the continuum
data consisted of 32 channels centred at 1.344~GHz with a total
bandwidth of 128~MHz, and the line data comprised 1024 channels centred at
1.420~GHz, with a total bandwidth of 4~MHz (a
channel separation of 0.83 \kms). All four Stokes parameters were
recorded in continuum, while only total-intensity data were
recorded in the \HI\ line. Amplitudes were
calibrated using the revised scale of Reynolds \shortcite{rey94},
assuming a flux density for PKS~B1934--638 of 15.0 and 14.9~Jy, at
1.344 and 1.420~GHz respectively (1~Jy $= 10^{-26}$ W m$^{-2}$
Hz$^{-1}$).  Phases were calibrated using PKS~B1549--790, which was
observed for approximately 3~min for each hour of observation.

\begin{table}
\caption{ATCA observations of \snra.}
\label{tab_observations}
\begin{tabular}{cccc} 
Date    & Array  & Maximum & Time on \\
	& Configuration & Baseline (m) & Source (h) \\ \hline
1996 Jan 19 & 0.75C & 750 & 12 \\
1996 Feb 02 & 0.75B & 765 & 11 \\
1996 Feb 24 &  1.5C & 1485 & 14 \\ \hline
\end{tabular}
\end{table}

Reduction and analysis were carried out using the Miriad \cite{stw95}
and Karma \cite{goo96} packages.  Data were edited and calibrated using
standard techniques \cite{sk97}. The calibrator PKS~B1549--790 (S$_{\rm
1.3\,GHz} = 5.4$~Jy) was confused by the source PKS~B1547--795 (S$_{\rm
1.3\,GHz} = 0.8$~Jy), 27 arcmin distant. Thus a point-source model for
the calibrator was not appropriate, and antenna gains were determined
by comparing the measured visibilities to those corresponding to a
model including both sources, in effect carrying out a self-calibration
of the calibrator field.

\subsection{Continuum data}
\label{sec_obs_continuum}

A total-intensity continuum image of a 65-arcmin square region was formed 
using multi-frequency synthesis \cite{sw94},
uniform weighting and a cell size of 5 arcsec.  The image was then
deconvolved using a maximum entropy algorithm \cite{gd78} with 500
iterations. To improve estimates of the antenna gains, three
iterations of amplitude and phase self-calibration were applied, with
a solution interval of 5~min. The resulting model was smoothed
with a Gaussian restoring beam, and the final image then corrected for
the mean primary beam response of the ATCA antennae.

Images were also formed in Stokes Q, U and V.  To minimise the effects
of bandwidth depolarization, 13 Q and U images were made, each using a
bandwidth of 8 MHz, covering the central 104~MHz of the band.  Stokes Q,
U and V images were then deconvolved using the CLEAN algorithm
\cite{cla80}, then restored and primary-beam corrected as for Stokes~I
above.  For each pair of Stokes Q and U images, a linear polarization
image, L,  was formed and corrected for non-Gaussian noise statistics
\cite{kbe86}.  A mean of the 13 L planes was then formed. This image
was then clipped at pixels where polarized emission fell below the 5-$\sigma$
level, and also where the total intensity was less than 5~$\sigma$.

\subsection{Line data}
\label{sec_obs_line}

Continuum emission was subtracted from the line data in the $u-v$
plane using the Miriad task UVLIN \cite{vc90,sau94}, and spectra
then smoothed to a velocity resolution of 3.3~\kms.  Data at
projected $u-v$ spacings shorter than 1~k$\lambda$ were discarded to
filter out broad-scale structure, and 
line images were then formed
in velocity planes ranging from --200 to +200 \kms\ (LSR).
The peak emission was faint, with no obvious sidelobes, and it was
consequently unnecessary to deconvolve the resulting images.

The \HI\ cube was then weighted by the continuum image. Absorption
spectra against continuum sources were produced by integrating over the
corresponding region of the cube, and then renormalising appropriately
to give units of fractional absorption.

The rms noise, $\sigma$, in each spectrum was estimated using the flux
in line-free channels. The brightness temperature of Galactic \HI\ in
this direction ($\sim100$~K; Caswell et al. 1975\nocite{cmr+75})
increases the thermal noise in channels containing line emission by a
factor of $\sim2$ (cf.\ Dickey 1997\nocite{dic97}). In
the spectra presented we thus show a threshold corresponding to 6~$\sigma$
as an indication of the believability of features.

\section{Results}
\label{sec_results}

\subsection{Total Intensity}
\label{sec_results_i}

\subsubsection{SNR~\snra}
\label{sec_results_i_snr}

Total-intensity images of SNR~\snra\ in Fig~\ref{fig_snr} show a
complicated structure similar to that seen with lower resolution
at 843~MHz \cite{wg96}. The SNR consists of multiple loops of emission,
brightest in the north-west, where there is a sharp outer
boundary.  While the western edge is particularly
well defined, the eastern perimeter is fainter and more diffuse.  

\begin{figure*}
\begin{minipage}{160mm}
\vspace{7.5cm}
\caption{Grey-scale and contour images of SNR~\snra\ at 1.3~GHz.
The grey-scale range is --1 to 20~mJy~beam$^{-1}$, while the
contour levels are 2.5, 5, 7.5, 10, 20, 30 and 
40~mJy~beam$^{-1}$. The FWHM of the Gaussian restoring beam is
shown at the lower right of each image.}
\label{fig_snr}
\end{minipage}
\end{figure*}

Running through the centre of the remnant is a peculiar, slightly
curved strip of emission, beginning beyond the remnant's northern
boundary and extending up, or close, to its southern perimeter.  This
structure shows no limb-brightening, and has an essentially uniform
appearance across its extent except for a possible fading to the
south.  Although appearing rather artificial, this feature is also seen
in the 843-MHz image made with the Molonglo Observatory Synthesis
Telescope (MOST; Whiteoak \& Green 1996\nocite{wg96}). Although both
are synthesis telescopes, the ATCA and the MOST produce quite
different artifacts \cite{rob91} and we conclude that this feature
corresponds to genuine emission.

Derived parameters for the SNR are given in Table~\ref{tab_snr}.  
The flux density
for the remnant and its uncertainty were determined by integrating the
emission within multiple polygons enclosing the remnant. A background
correction was made and the rms noise in the image measured by
similarly integrating over nearby source-free regions.

\begin{table}
\caption{Observational and derived parameters for \snra.}
\label{tab_snr}
\begin{tabular}{lc} \hline 
Resolution & $24\farcs3 \times 22\farcs4$, PA 14$^{\circ}$ \\
rms noise in image & 170 (Stokes I) \\
\hspace{1cm} ($\mu$Jy~beam$^{-1}$) &  40 (Stokes V) \\
Geometric centre ($\alpha$, $\delta$; J2000) & 
$11^{\rm h}58^{\rm m}25^{\rm s}$ --62\degr36\arcmin \\
Geometric centre ($l$, $b$) & 296\fdg87 --00\fdg35 \\
Diameter (arcmin) &  $18 \times 12$ \\
Flux density at 1.3~GHz (Jy) & $7.0\pm0.3$\\
Spectral index & $-0.65\pm0.08$ \\ \hline
\end{tabular}
\end{table}

We compute a spectral index for
SNR~\snra\ using the total flux density
measurements shown in Table~\ref{tab_spectrum}. Fitting a power law of the 
form $S_\nu \propto \nu^{\alpha}$ to the data, we obtain a
spectral index $\alpha = -0.65 \pm 0.08$,
consistent with earlier determinations \cite{lv72,ccc75}.

\begin{table}
\caption{Flux density measurements of SNR~\snra. Where not specified by
the authors,
errors have been taken to be 10 per cent or 1~Jy, whichever is larger.}
\label{tab_spectrum}
\begin{tabular}{clcc} 
$\nu$ (GHz) & Telescope      & S$_\nu$ (Jy)       & Reference \\ \hline
0.408       & Mills Cross    & $15.0\pm1.5$ &     1 \\      
0.843       & MOST           & $9.2\pm0.9$  &     2 \\     
1.344       & ATCA           & $7.0\pm0.3$  &  this paper \\
1.420       & Parkes 64-m     & $6\pm1$      &  1,3 \\ 
2.650       & Parkes 64-m     & $4\pm1$      &    4 \\
5.000       & Parkes 64-m     & $3.2\pm1$    &    5 \\ \hline
\end{tabular}
\\
\footnotesize
(1)~Large \& Vaughan~\shortcite{lv72} 
(2)~Whiteoak \& Green~\shortcite{wg96} 
(3)~Hill~\shortcite{hil68}
(4)~Thomas \& Day~\shortcite{td69} 
(5)~Caswell, Clark \& Crawford~\shortcite{ccc75} 
\end{table}

\subsubsection{Other sources in the field}
\label{sec_results_i_other}

An image of the entire field is shown in
Fig~\ref{fig_primary_beam}.  The properties of four sources of note
are summarised in Table~\ref{tab_sourcelist}.

\begin{figure*}
\begin{minipage}{140mm}
\vspace{16cm}
\caption{A total-intensity image of the field surrounding
SNR~\snra. In order to give uniform noise across the image,
the image has not been corrected for the ATCA primary beam
response. Sources listed in Table~\ref{tab_sourcelist} are
indicated, as well as the two regions of the SNR against which
\HI\ absorption was obtained. The line to the north of the SNR
represents the Galactic Plane.}
\label{fig_primary_beam}
\end{minipage}
\end{figure*}

Source 1 can be identified with the 400-ms pulsar
\mbox{J1157--6224} \cite{lvw69b,smd93}. 
We measure an integrated flux density for the pulsar of $9.7\pm0.3$~mJy,
agreeing with previous measurements 
\cite{jkww96}.  

\begin{table*}
\begin{minipage}{150mm}
\caption{Selected sources in the vicinity of SNR~\snra. All sources are
unresolv
ed.}
\label{tab_sourcelist}
\begin{tabular}{cccccll} 
Source  & \multicolumn{2}{c}{Position}   & 
$S_{\rm 1.3\,GHz}$ & Spectral index$^a$  & Other name \\
& Equatorial (J2000) & Galactic 
& (Jy) & ($\alpha$, $S_{\nu} \propto \nu^{\alpha}$) & & \\ \hline
  1     & 11$^{\rm h}$57$^{\rm m}$15$^{\rm s}$ --62\degr24\arcmin50\arcsec &
  296\fdg70 --00\fdg20 & 0.01 & --2.6 & PSR~J1157--6224 \\
  2     & 11$^{\rm h}$57$^{\rm m}$20$^{\rm s}$ --62\degr43\arcmin50\arcsec & 
  296\fdg78 --00\fdg51 & 0.08 & --0.5 &  \\
  3     & 11$^{\rm h}$59$^{\rm m}$32$^{\rm s}$
  --62\degr07\arcmin19\arcsec & 
   296\fdg90 --00\fdg14 & 0.62 & --0.9 & PMN~J1159--6207 \\
  4     & 12$^{\rm h}$01$^{\rm m}$21$^{\rm s}$
  --62\degr36\arcmin44\arcsec & 297\fdg21 --00\fdg30 & 0.11 & --0.6 &  \\ \hline
\end{tabular}
	
$^a$ Calculated between 1.344~GHz (this paper) and 843~MHz (A.J.\ Green 
et al., in preparation).
		 
\end{minipage} 
\end{table*}

\subsection{Polarization}
\label{sec_results_pol}

\subsubsection{PSR~J1157--6224}
\label{sec_results_pol_psr}

As a test of the polarimetric quality of our data, we first
consider PSR~J1157--6224, whose polarization properties we
can compare with existing data.

PSR~J1157--6224 is significantly linearly polarized: we measure a
linear polarization L~$=2.5\pm0.3$~mJy 
($26\pm4$ per cent), agreeing well with
previous measurements \cite{mhma78,mhm80,vdhm97}. 

The ATCA generates instrumental linear
polarization proportional to the strength of a source
in total intensity. This response is negligible at
the phase centre, but becomes significant towards
the edges of the field. The primary-beam corrected
linear polarization for the pulsar caused by this effect
is of the order of 0.2 per cent of the total intensity
and makes a minor contribution to the value obtained.

\begin{figure}
\centerline{\psfig{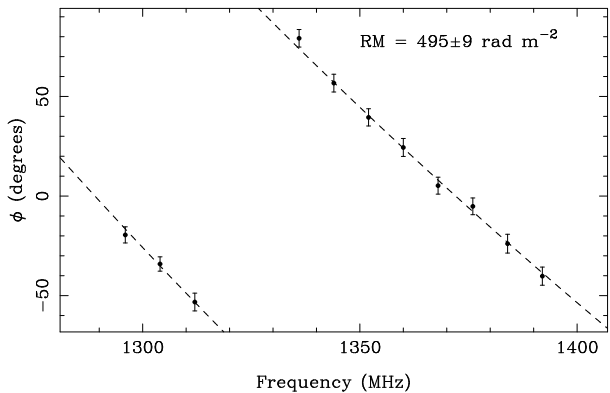}}
\caption{A plot of the Faraday rotation across the band for
PSR~J1157--6224. The plotted points correspond to position angles
measured at 8-MHz intervals (gaps at 1320 and 1328~MHz correspond
to bad data), while the line corresponds to the best fitting
curve of the form $\phi = \phi_0 + {\rm RM}\, c^2/\nu^2$.}
\label{fig_psr_rm}
\end{figure}

The multi-frequency capability of the ATCA allows us to measure the
variation in position angle, $\phi$, of the pulsar's linearly polarized
emission across the observing bandwidth, as shown in
Fig~\ref{fig_psr_rm}.  This can be attributed to 
Faraday rotation, and we can hence derive a rotation
measure for the pulsar.  Fitting a curve with functional form $\phi =
\phi_0 + {\rm RM}\, c^2/\nu^2$ to the data gives a rotation measure
RM~$= 495\pm9$~rad~m$^{-2}$.  Within the uncertainties, this agrees
with the published
value of $508\pm5$~rad~m$^{-2}$ \cite{tml93}.

PSR~J1157--6224 is left-hand circularly polarized with a flux density
$1.2\pm0.2$~mJy ($12\pm2$ per cent of the total). This is again
consistent with earlier work.

In summary, we conclude that the observed polarization properties of
PSR~J1157--6224 agree well with previous measurements.

\subsubsection{SNR~\snra}
\label{sec_results_pol_snr}

\begin{figure}
\centerline{\psfig{file=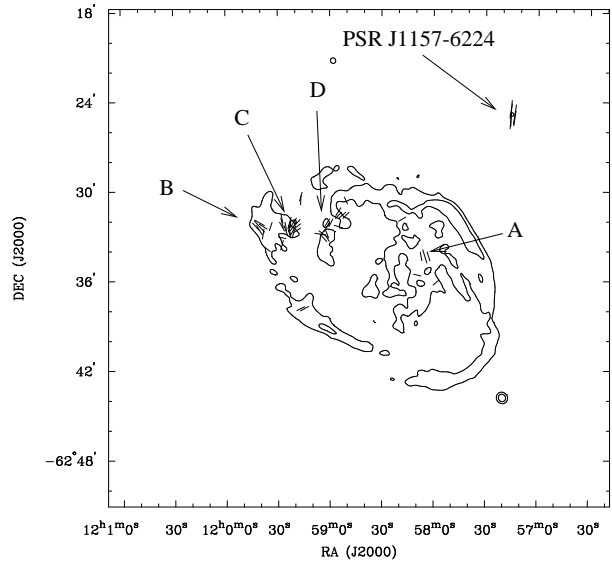,width=8cm,angle=270}}
\caption{Linearly polarized emission from \snra. The orientation
of vectors indicate the position angle of the electric
field in an 8-MHz channel centred on 1376~MHz. Lengths of vectors
are proportional to the surface brightness in linear polarization at that
position, the longest vector corresponding to
${\rm L}=2.5$~mJy~beam$^{-1}$. Contours represent total intensity, and
are drawn at 5 and 20 mJy~beam$^{-1}$.
The labelled regions correspond to the plots in
Fig~\ref{fig_snr_rm}.}
\label{fig_snr_pol}
\end{figure}

No circular polarization is detected from SNR~\snra.  An image of the
linear polarization from the remnant is shown in
Fig~\ref{fig_snr_pol}. We estimate a lower limit on the linearly
polarized flux density from the remnant of $35\pm5$~mJy, a fractional
polarization of 0.5~per cent (instrumental polarization contributes
less than 0.1~per cent).  Polarization from the SNR occurs in a
few discrete cells, predominantly near the north-eastern edge of the
SNR (note that region `A' is not coincident with the compact knot in
this region of the SNR). Within a cell there appears to be some
correlation in position angle but adjacent cells have quite different
orientations.  There is no correlation between total and polarized
intensity over the SNR.  In fact the brightest regions of the SNR,
namely the north-west, south-west and south edges, have no significant
polarized emission.

\begin{figure}
\centerline{\psfig{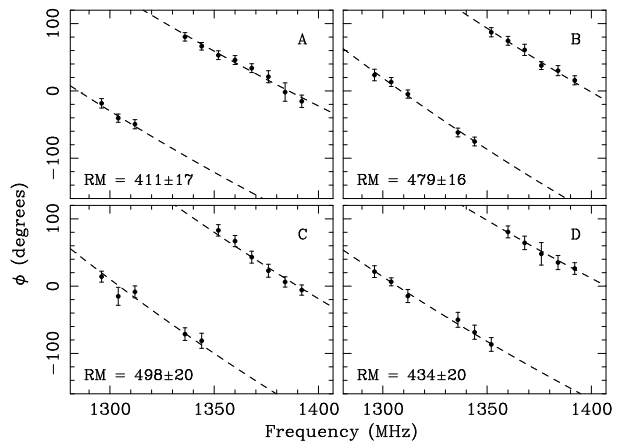}}
\caption{Faraday rotation and rotation measures (in rad~m$^{-2}$) for
four regions  marked in  Fig~\ref{fig_snr_pol}.  Properties of the
plots are as for Fig~\ref{fig_psr_rm}.}
\label{fig_snr_rm}
\end{figure}

In Fig~\ref{fig_snr_rm} we show the position angle as a function of
frequency for four regions of significant linear polarization.  As for
PSR~J1157--6224, a change in position angle of the electric field
vectors is seen across the band.  
Averaging over regions of sufficiently strong polarization, we obtain a
mean RM for the remnant of 430~rad~m$^{-2}$, with a scatter of
40~rad~m$^{-2}$.

In theory, one can use the rotation measure to determine the
intrinsic position angle of the polarized radiation, $\phi_0$.
However, the accuracy, $\delta\phi_0$, of this measurement is limited by
the uncertainty in RM, $\delta$(RM). Typical values of
$\delta$(RM) across the SNR are 20--30~rad~m$^{-2}$, 
corresponding to
$\delta\phi_0 \approx \delta({\rm RM})\,c^2/\nu^2
\approx 1$. Thus we cannot determine intrinsic position angles,
effectively because of the narrow bandwidth across which we have made our
measurements.

\subsubsection{Other sources}
\label{sec_results_pol_other}

After correcting for instrumental polarization, sources 2, 3 and 4
are 0.2, 2.7 and 1.4 per cent linearly polarized respectively.
No circular polarization is detected from any of these sources.

\subsection{\HI\ line}
\label{sec_results_line}

In the following discussion we establish upper and lower limits on 
the systemic velocity of SNR~\snra\ by measuring the
\HI\ velocities of absorbing clouds along the line of sight and
combining these with an appropriate model for Galactic rotation to derive
a kinematic distance to the remnant (see Burton 1988\nocite{bur88b} for
a review). We compare absorption against the remnant to adjacent
emission and absorption spectra; the former is obtained from nearby
regions devoid of continuum emission, while the latter is seen
towards sources 2 and~4.

We define the lower limit on the systemic velocity, $V_L$, to be that
corresponding to the most distant absorption feature seen against the
SNR, and the upper limit, $V_U$, to be the nearest strong emission
feature in the region not seen in absorption (e.g.\ Frail \& Weisberg
1990\nocite{fw90}).  We adopt an uncertainty of $\pm7$~\kms\ in $V_U$
and $V_L$, representative of the random motion of \HI\ clouds
\cite{sra+82,bc84};
uncertainties associated with estimating
velocities from our spectra are somewhat smaller than this.  When
translating velocities into distances, we adopt
standard IAU parameters \cite{klb86} for the solar orbital velocity
($\Theta_0 = 220$~\kms) and the distance to the Galactic Centre ($R_0 =
8.5$~kpc).

\begin{figure}
\centerline{\psfig{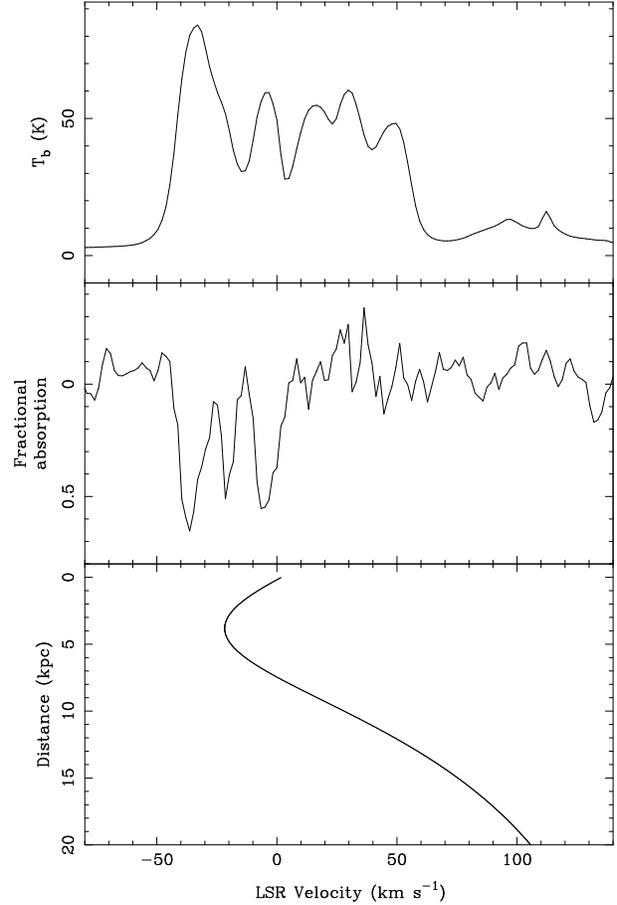}}
\caption{\HI\ emission and absorption towards PSR~J1157--6224, and the
variation of velocity with distance from the sun in this direction
(Johnston et al. 1996).}
\label{fig_hi_psr}
\end{figure}

The lack of short $u-v$ spacings in our ATCA observations means that it
is difficult to produce an emission spectrum from these data. However,
we can compare ATCA absorption spectra to the 
emission spectrum seen
towards PSR~J1157--6224 \cite{jkww96}, only 7 arcmin from the
brightest part of the SNR. This spectrum is shown in Fig~\ref{fig_hi_psr}, 
together with the single-dish absorption towards the pulsar
and the rotation curve in this direction. The emission spectrum contains
strong ($T_b > 35$~K) features at --35, --5, +15, +30 and +50
\kms. Similar emission profiles are seen in the direction of the
nearby \HII\ regions G298.2--00.3 and G298.9--00.4 \cite{grbm72} and
in the \HI\ survey of Kerr et al. \shortcite{kbjk86}.

\subsubsection{Point sources}
\label{sec_results_line_point}

\begin{figure*}
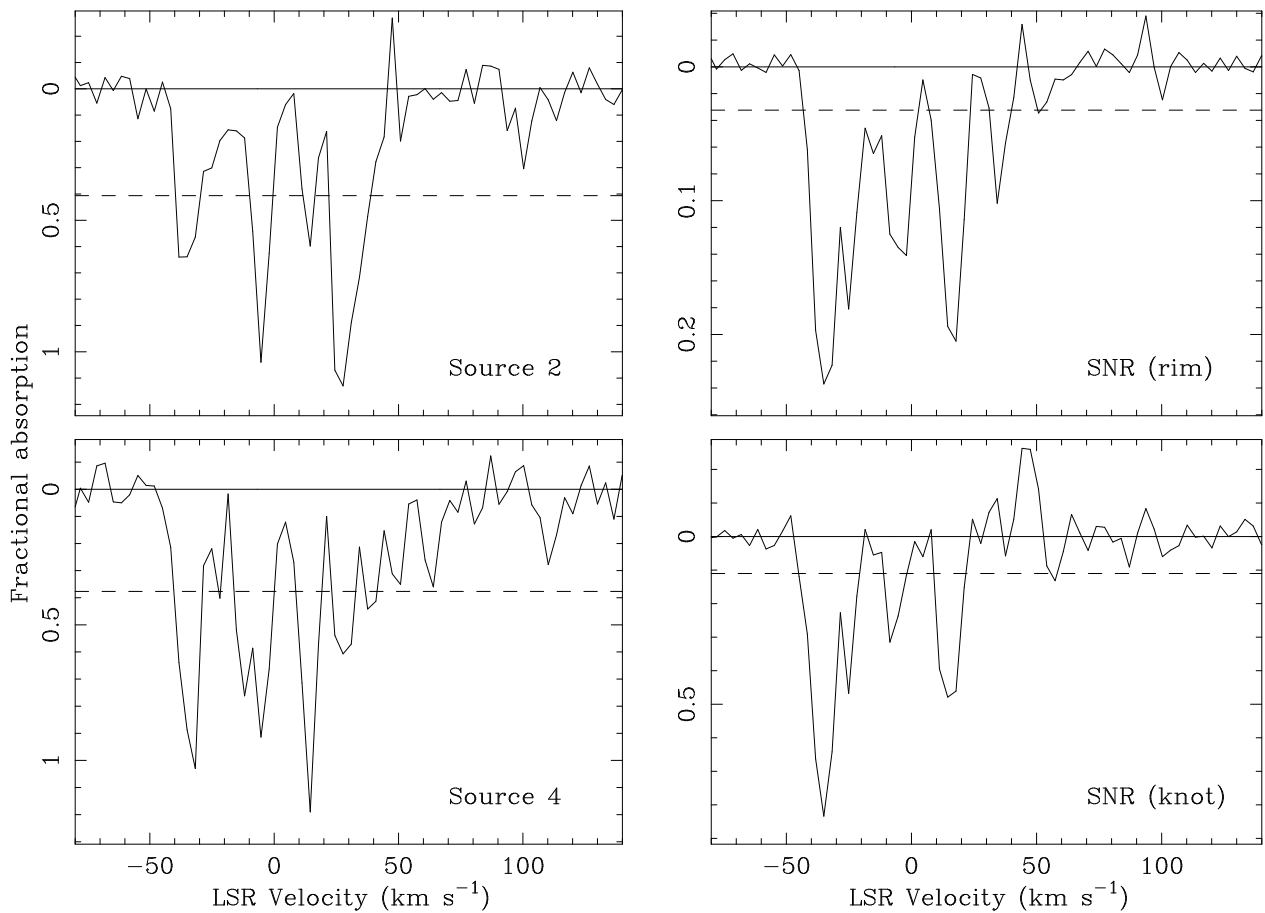

\begin{minipage}{160mm}
\centerline{\psfig{file=figure7a.eps,height=12cm}
\hspace{5mm}
\psfig{file=figure7b.eps,height=12cm}}
\caption{\HI\ absorption spectra towards sources 2 and 4 and towards
the regions of SNR~\snra\ indicated in Fig~\ref{fig_primary_beam}. 
The dashed line represents absorption at the 6-$\sigma$ level, where $\sigma$
is calculated from the emission in line-free channels.}
\label{fig_hi}
\end{minipage}
\end{figure*}

Absorption was measured against the nearby sources 2 and 4, with
spectra shown in Fig~\ref{fig_hi}.
The absorption spectrum obtained towards source 3 was noisy and not
considered useful.

The most negative velocity absorption feature seen in the spectrum
of source 2 is at --35~\kms, agreeing well with the measured tangent
velocity in this direction \cite{kbjk86,mal95}.
Significant absorption is also seen at --5, +15 and +30~\kms.
A weak feature at +100~\kms\ is of low significance.

Source 4 shows similar structure to source 2 at negative velocities.
At positive velocities, strong absorption is seen at +15~\kms, and 
a forest of weaker features is visible at +30, +40, +50, +65 and +110~\kms.
Although each feature individually is of marginal significance,
together they suggest true absorption rather than random fluctuations.

\subsubsection{SNR~\snra}
\label{sec_results_line_snr}

Spectra of reasonable signal-to-noise were seen against the two regions of
the SNR indicated in Fig~\ref{fig_primary_beam}:  the bright rim along
the SNR's north-west edge, and a nearby knot in the interior. These
spectra are shown in Fig~\ref{fig_hi}.

Against the rim of the SNR, absorption seen at negative velocities
corresponds to emission seen towards the pulsar and absorption towards
sources 2 and 4.  At positive velocities, clear absorption is seen at
+15~\kms\ and a weaker feature at +35~\kms. Beyond this the spectrum is
consistent with noise.  Absorption towards the knot at negative
velocities is similar to that towards the rim.  Strong absorption is
also seen at +15~\kms but not beyond this point.

The weak feature seen against the rim at +35~\kms\ matches the
+30-\kms\ feature clearly seen towards source 2 only approximately.
While this difference in velocity is within the uncertainties
associated with \HI, emission towards the pulsar, much closer to the
rim than source 2, also shows a feature peaking at +30~\kms. This seems
to indicate a genuine discrepancy in velocity between the
+35-\kms\ feature towards the SNR and +30-\kms\ absorption towards
source 2.  Furthermore, while the two SNR spectra are otherwise
similar the +35-\kms\ feature is not seen against the knot, despite
its proximity to the rim.  One could argue that the knot is an
unrelated foreground source but we think it more likely that the
+35-\kms\ feature represents fluctuations in emission rather than true
absorption, and assign lower and upper limits to the remnant of $V_L =
+15$~\kms and $V_U = +30$~\kms\ respectively.  Given an uncertainty of
$\pm7$~\kms, we can assign an overall systemic velocity to the SNR of
$V_{\rm SNR} = 23\pm7$~\kms.

\section{Discussion}
\label{sec_discuss}

\subsection{SNR~\snra}
\label{sec_discuss_snr}

\subsubsection{Physical Parameters}
\label{sec_discuss_snr_params}

Adopting the best fitting model for Galactic rotation 
of Fich, Blitz \& Stark \shortcite{fbs89}, we use
the velocity determined for the SNR in
Section~\ref{sec_results_line_snr} to derive a kinematic distance
of $9.6\pm0.6$~kpc, putting it on the far side of the Carina
arm. This is consistent with Hwang \& Markert's \shortcite{hm94}
estimate based on the lack of X-ray emission.  A distance of 9.7~kpc
has also been estimated from the $\Sigma - D$ relation \cite{cb83b}.
However, given the large uncertainties in this method \cite{gre84}, this
agreement is certainly fortuitous.

\begin{figure}
\centerline{\psfig{file=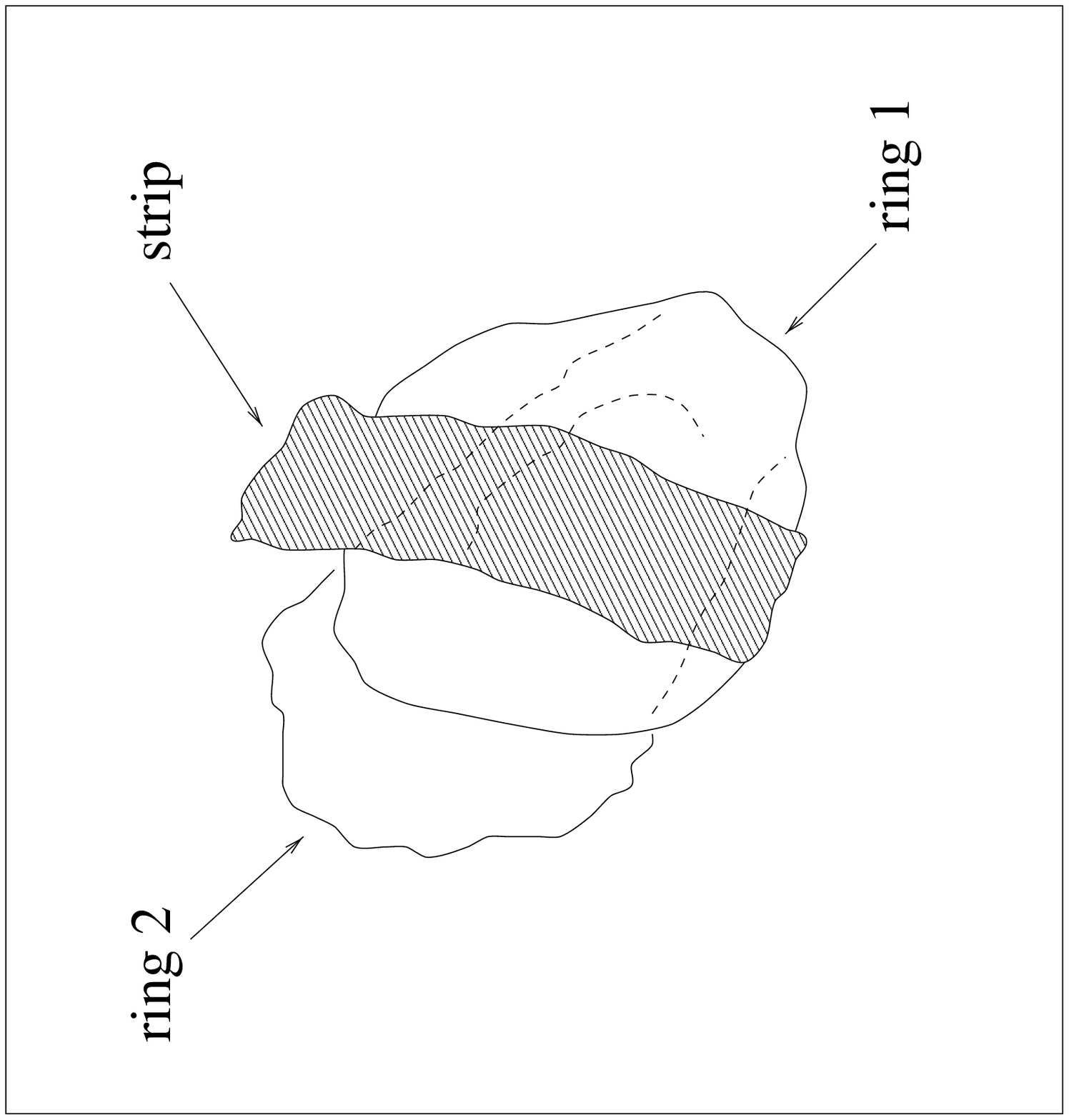,width=8cm,angle=270}}
\caption{A sketch of the morphology of SNR~\snra, showing some
pertinent features.}
\label{fig_cartoon}
\end{figure}

Using the brightest regions (`ring 1' -- see 
Fig~\ref{fig_cartoon})
to define a single SNR shell, we find a
radius for \snra\ of $17\pm1$~pc.  An upper limit on the remnant's age
can consequently be obtained by assuming that it has expanded into the
ambient ISM.  For an ambient density $n_0$~cm$^{-3}$, the mass swept
up by the SNR is $\sim500n_0\, M_{\sun}$.  Typical values of $n_0$
then imply that the ejecta have swept up many times their own mass and
that the remnant is no longer freely expanding. If we assume that the
remnant is in the adiabatic (Sedov--Taylor) phase, we can derive an
age $t_{\rm SNR} = (22\pm3) \left(n_0/E_{51} \right)^{1/2} \times
10^3$~yr, where $E_{51}$ is the kinetic energy of the explosion in
units of $10^{51}$~erg. Frail, Goss \& Whiteoak \shortcite{fgw94}
suggest typical values $n_0 = 0.2$ and \mbox{$E_{51} = 1$}, implying 
that $t_{\rm SNR} = (10\pm2) \times 10^3$~yr.

We derive a lower limit on the remnant's age by assuming free
expansion, as would occur if the remnant is propagating through the
main-sequence wind bubble of a massive progenitor \cite{cl89,tbfr90}.
Free expansion at $(0.3-1.0) \times 10^4$~\kms\ gives an age in the range
1600--6000~yr.

\subsubsection{Polarization}
\label{sec_discuss_snr_pol}

The rotation measure of 430~rad~m$^{-2}$ derived for the SNR in
Section~\ref{sec_results_pol_snr} 
differs from the value of \mbox{$\sim100$~rad~m$^{-2}$} measured by Dickel
\& Milne \shortcite{dm76} over the north-east part of the remnant.
Given that we obtain polarization data for PSR~J1157--6224
consistent with earlier measurements, we have confidence in our results.
In any correction for Faraday rotation, there is always an
ambiguity of $n\pi$ ($n = 0,\pm1,\pm2,\ldots$).
The discrepancy in rotation measures can be accounted for if Dickel \&
Milne's\nocite{dm76} result (obtained using data at 2.7 and 5~GHz) is
out by a half-turn, causing them to underestimate the rotation measure
by 360~rad~m$^{-2}$. It is unlikely that there is any further ambiguity
in the data: at the 8-MHz frequency resolution of the ATCA observations,
a further half-turn would increase the RM by $\sim5000$~rad~m$^{-2}$.

Although the theoretical fractional polarization of synchrotron
radiation is 70 per cent, this SNR is quite typical in having a much
lower value, indicating that the remnant has been depolarized.
We can immediately rule out bandwidth depolarization as a possible cause
since, by imaging the polarized emission in several frequency planes
before combining, we have reduced Faraday rotation across
each channel to $<20\degr$.

The cell structure of the polarized emission, with significant changes
in position angle between adjacent regions, suggests that the SNR has
been beam depolarized, 
i.e.\ that large changes in position angle are occurring on scales
smaller than the resolution element.  This could be a result of
differential Faraday rotation in the ISM, or may correspond to
intrinsic structure in the SNR's magnetic field. The former possibility
is certainly reasonable, given that it requires fluctuations in RM
within a beam of only $\sim20$~rad~m$^{-2}$, comparable to the
variation in RM seen across the source. However, to determine
conclusively which mechanism applies will require observations at
higher frequencies, where effects caused by Faraday rotation are minimal.

Using the Galactic electron density distribution of Taylor \&
Cordes \shortcite{tc93}, we compute an electron column 
density along the line of sight to the SNR of 
260--330~pc~cm$^{-3}$.  Assuming that Faraday rotation
internal to the SNR makes only a small contribution to the total
rotation measure, 
we compute an averaged line-of-sight interstellar magnetic field of
$1.7\pm0.3$~$\mu$G, directed towards us.

\subsubsection{Morphology}
\label{sec_discuss_snr_morph}

The morphology of SNR~\snra\ is not that seen in a typical shell SNR,
and we suggest three possible explanations to account for its
complicated and multi-shelled appearance. First, the SNR could be two
separate objects. Second, it may have a double-ringed morphology
corresponding to an underlying biannular symmetry in the progenitor
system, as proposed by Manchester \shortcite{man87}. Finally,
\snra\ may owe its appearance to the surrounding ISM. We now consider
these various possibilities.

Given that the lifetime of a supernova progenitor exceeds the
observable lifetime of a SNR by 2--3 orders of magnitude, it is highly
unlikely that we are seeing two SNRs from the same stellar assocation
in temporal co-existence.  It is possible that we are seeing
two SNRs lying at different distances but along the same line of
sight, but we think that this too is unlikely as it would
require the centres of the only two SNRs in the field to virtually
coincide.  Further circumstantial evidence against \snra's being
multiple SNRs is that \HI\ absorption from the rim and the knot give
consistent systemic velocities, and that the rotation measure is
roughly uniform across the SNR.
Thus the possibility of multiple SNRs cannot be ruled out by
our data, but is unlikely.

It is now well-established that the slow wind from the red supergiant
(RSG) phase of massive stars can possess axial symmetry (Blondin,
Lundqvist \& Chevalier 1996\nocite{blc96} and references therein).
Manchester \shortcite{man87} proposed that expansion of a SNR into such
a medium can produce a biannular, or double-ringed morphology, similar
to models proposed for planetary nebulae \cite{kw85,bal87}. Theory
\cite{its92,blc96} and observations \cite{gms+97} both support the
argument that this axisymmetric RSG wind can indeed affect the shape
and morphology of a remnant expanding through it.  Typical parameters
for the RSG phase are a wind speed $v_{\rm RSG} = 20$~\kms, a mass loss
rate $\dot{M}_{\rm RSG} = 10^{-7} - 10^{-5}$~M$_{\sun}$~yr$^{-1}$ and
a life-time $t_{\rm RSG} = 3 \times 10^5$~yr \cite{smi94}. This
generates a potentially asymmetric medium around a supernova of
radius~$<$5~pc.  Assuming such an environment for SNR \mbox{\snra}, we
find that the remnant, with a radius of $\sim$17~pc, has expanded well
beyond this region and into the main-sequence wind of the progenitor.
In the spherically symmetric case, this wind is predominantly hot,
isobaric and consequently isotropic \cite{wmc+77}, and we expect a SNR
propagating through it to rapidly become spherical \cite{bb82}.  For such a
SNR to retain the brightness distribution induced in it by the RSG wind
requires a mechanism by which shock acceleration
preferentially occurs in the biannular emitting regions \cite{man87}.
A more likely explanation is the presence of asymmetry in the main
sequence wind: for example, the effect of rotation in the progenitor
can concentrate such a wind along the equatorial plane
\cite{bc93,ocb94}. 
To summarise, SNR~\snra\ is well beyond the size at which biannular
structure could be imprinted upon it by asymmetric mass loss in a RSG
progenitor wind. If its appearance is caused by the structure of
circumstellar material, then we require either the SNR to have somehow
retained a biannular morphology while expanding into an isotropic
medium, or the progenitor's main-sequence wind to have possessed
axial symmetry.

Manchester \shortcite{man87} also suggests that a biannular SNR could be
formed by the preferential distribution of ejecta along the polar
axis.  However, evidence suggests that asymmetric ejecta do not produce
an unusually shaped remnant \cite{bb82,blc96,gms+97}.

A further means of producing biannular structure is through 
jets from a central pulsar or binary system, which then hit the
surface of the expanding remnant \cite{man87}. Our radio observations
of SNR~\snra\ show no evidence for a centrally located point source
either in total or polarized intensity, nor for outflows from such a
source.  The X--ray observations also show no central emitting region
which might correspond to a central star or nebula.  Apart from
J1157--6224, no pulsar has been detected in or near this remnant
\cite{mdt85,kmj+96}.  We conclude that there is no evidence
for a central source or related outflows in SNR~\snra.

We now consider the possibility that the SNR is interacting with the
ambient medium. In order to propose a simple model of the SNR's appearance,
we denote the main features in the SNR as follows, illustrated
schematically in cartoon form in Fig~\ref{fig_cartoon}:

\begin{enumerate}
\item{} a bright, sharp-edged, and almost complete ring of emission
in the south-west (`ring 1'). In the absence of other structure, this
component would appear as a typical shell SNR;
\item{} a rectangular strip of emisssion running north-south through
the centre of the SNR.  This strip is not a typical SNR feature -- it
is almost linear, is uniform in brightness over its extent, with
projected dimensions of $\sim40$~pc north-south and $\sim13$~pc east-west
(at a distance of 10~kpc);
\item{} a faint, ragged partial ring (`ring 2'), extending beyond ring~1
to the north-east and possibly connecting with filaments in the south.
\end{enumerate}

Ring 1 seems compatible with a SNR expanding
into a homogeneous medium. We suggest that the strip corresponds to a
tunnel or tube of significantly lower density.  This is of appropriate
dimensions to be an old SNR or stellar-wind bubble, its elongated
appearance caused by tension in external \cite{sn92,gae98} or internal
\cite{cs74,lprv89} magnetic field lines.  On encountering such a region,
an expanding shock will re-energise it, rapidly propagating both across
the tube and up and down its length. As it does so, electrons in the walls
are shock-accelerated and emit synchrotron radiation \cite{plr87}, giving
a filled, linear structure as observed here. 

Ring 2 is fainter than ring 1, extends beyond the latter's boundary,
and has a more poorly defined outer edge. This suggests that it
represents a region of break-out into an adjacent cavity, perhaps
created by another massive star. In such a case the shock will rapidly
expand to take on the dimensions of the region, creating the multiple
shell appearance observed \cite{bs86}

It has been argued that the morphologies of SNRs~G166.0+04.3 (VRO
42.05.01) and G350.0--02.0 both result from a supernova shock
re-energising a tube or tunnel and then propagating on to the other
side to form a second shell \cite{plr87,gae98}. In both these SNRs, the
three components are all approximately in the plane of the sky. We
suggest that SNR~\snra\ may represent a similar situation, but in this
case where the explosion site, the low-density tunnel and the second
shell all lie roughly along the line of sight.

\subsection{PSR~J1157--6224}
\label{sec_discuss_psr}

\subsubsection{Physical Properties}
\label{sec_discuss_psr_params}

The similarity in the rotation measures of the SNR and the pulsar
suggests that they are at roughly the same distance.  The poor
signal-to-noise in the absorption spectrum shown in
Fig~\ref{fig_hi_psr} suggests that while the Johnston et
al.\ \shortcite{jkww96} lower limit on the kinematic distance of 4~kpc
(inferred from absorption out to the tangent point) is valid, their
upper limit of 9~kpc, based on lack of absorption at positive
velocities, can be questioned. Indeed an early \HI\ absorption spectrum
towards the pulsar suggested absorption out to +30~\kms\ \cite{am76}.
Putting the pulsar at a similar distance to the SNR gives
agreement with the dispersion measure distance derived from the model
of Taylor \& Cordes \shortcite{tc93}. Thus we propose a distance for
the pulsar of 10~kpc.

\subsubsection{Polarization}
\label{sec_discuss_psr_pol_psr}

A Stokes~V image of the entire field is shown in
Fig~\ref{fig_stokes_v}.  Because the ATCA has linear feeds,
the instrumental response in Stokes~V is negligible.  Apart
from the pulsar, the image is devoid of emission, with a noise
(40~$\mu$Jy~beam$^{-1}$) commensurate with that expected from the
radiometer equation alone.

\begin{figure}
\vspace{8cm}
\caption{An image of the field in circular
polarization.  PSR~J1157--6224 is the only detectable
source.}
\label{fig_stokes_v}
\end{figure}

The signal-to-noise of the pulsar in Stokes~V ($\sim$30~$\sigma$) is
comparable to that in total intensity, yet the circularly polarized
signal-to-noise of all other sources in the field is zero.  Indeed
pulsars are generally $\sim$10 per cent circularly polarized
(e.g.\ Qiao et al.  1995\nocite{qmlg95}), while other sources tend not
to emit in circular polarization.  We therefore suggest that
examination of Stokes~V images may be a fruitful method of finding new
pulsars, particularly those which are too weak, too dispersed, too
scattered or have a period too short or too long to be
found by traditional searches for pulsed emission.

\subsection{Other Sources}
\label{sec_discuss_other}

Absorption against source 2 (Fig~\ref{fig_hi}) is not seen at +50~\kms,
where emission is seen in Fig~\ref{fig_hi_psr}.  However, at this
distance absorbing clouds are of small angular scale and may not lie
along this line of sight. This source has no counterpart in {\em IRAS}
60-$\mu$m emission, which, combined with its intermediate spectral
index and low linear 
and circular polarization, is consistent with its
being a background radio galaxy.  Source 4's \HI\ absorption suggests
that it too is extragalactic, consistent, as for source 2, with its
spectral, infrared and polarization properties.  While no absorption
was obtainable against source 3, the available evidence is also
consistent with an extragalactic origin.

\subsection{An association between SNR~\snra\ and PSR~J1157--6224?}
\label{sec_discuss_assoc}

An association between SNR~\snra\ and PSR~J1157--6224 was
originally proposed by Large \& Vaughan \shortcite{lv72}, primarily
based on the spatial coincidence of the two objects.  This association
has subsequently been regarded as unlikely on the basis of the pulsar's
large timing age $\tau_c = 1.6 \times 10^6$~yr \cite{gj95c,jkww96},
clearly incompatible with the age for the SNR derived above.  However,
the timing age of a pulsar is only an upper limit on the true age,
since it assumes that the initial period of the pulsar was much smaller
than the present value. If a pulsar is born spinning slowly
\cite{vn81,ec89,no90}, it can be much younger than $\tau_c$. Given
the evidence presented above that the distances of the SNR and the
pulsar are compatible, we now reconsider the likelihood of an
association.

If the pulsar is as young as $t_{\rm SNR} \la 10^4$~yr,
its position well outside the remnant argues strongly against a
physical association.  The pulsar is removed from the remnant's centre
by approximately double the SNR radius. Such a large displacement
is statistically improbable for an association
of this age \cite{gj95b}.  The implied projected velocity is $>3500$~\kms, a
speed similar to that claimed by Caraveo \shortcite{car93} in order to
associate PSR~J1614--5047 with G332.4+00.1, but well in excess of
typical pulsar velocities \cite{ll94,fgw94}.  We note, however, that the
implied proper motion of 80~mas~yr$^{-1}$ cannot be ruled out;
comparison of timing positions derived by Newton, Manchester \&
Cooke \shortcite{nmc81} and Siegmen et al. \shortcite{smd93} gives
an upper limit on the proper motion of 150~mas~yr$^{-1}$.

An alternative is that the remnant is much older than the age
calculated above. If we suppose that the remnant has an age of
$10^5$~yr, then the pulsar's inferred projected velocity ($\sim
400$~\kms) becomes quite reasonable, as does its position \cite{gj95a}.
However, this requires that $n_0/E_{51} \sim 25$, an usually high value
(cf.\ Frail et al.  1994\nocite{fgw94}).  The high pulsar velocity and
ambient density required for an association would be expected to
generate a wind nebula or bow shock trailing out behind the pulsar
(e.g.\ Frail \& Scharringhausen 1997\nocite{fs97}).  No such structure
is visible around the pulsar in our radio image, and comparison of the
integrated flux density (Table~\ref{tab_sourcelist}) with the pulsed
flux \cite{jkww96} puts an upper limit on emission from any unresolved
nebula of $\la0.1$~mJy.
Furthermore, the pulsar would still have to be born spinning slowly:
for a braking index $n=3$ \cite{mt77}, $t = 10^5$~yr requires an
initial period $P_0 = 388$~ms (compared with a current period of 400~ms).

Thus, we conclude that, apart from an agreement in
distance, there is no evidence in favour of a physical relationship
between SNR \snra\ and PSR~J1157--6224.  Most likely their spatial proximity
is by chance and the pulsar is a much older object.

\section{Conclusion}
\label{sec_conclusion}

We have presented 1.3-GHz continuum and \HI\ observations of
SNR~\snra.  These data (resolution $\sim$20 arcsec) represent the
highest resolution at which this SNR has been observed.  The remnant is
weakly polarized at a lower limit of 0.5 per cent. We attribute this to
beam depolarization, possibly a result of differential Faraday rotation
in the ISM.  The  ATCA's capacity to measure many channels across the
continuum band allows a measurement to be made of the change in
polarization position angle as a function of frequency. We consequently
derive a rotation measure towards the SNR of $430$~rad~m$^{-2}$. This
disagrees with the earlier results of Dickel \& Milne \shortcite{dm76},
and we argue that these authors incorrectly accounted for an ambiguity
in the rotation.

\HI\ absorption towards the SNR gives lower and upper limits on its
systemic velocity of +15 and +30~\kms\ respectively, corresponding to
a distance $9.6\pm0.6$~pc.  We consequently estimate an age for the
remnant of $(2-10) \times 10^3$~yr, and a  mean interstellar magnetic
field along the line of sight of $1.7\pm0.3$~$\mu$G.

SNR~\snra\ seems consistent with its being the remnant of a single
explosion.  Two explanations for its morphology seem possible. One
is that the remnant has a biannular appearance induced by
axial symmetry in the progenitor wind.  However, the SNR is much larger
than the RSG wind of a SN progenitor, and it must either be able to
`remember' its environment from an earlier stage of evolution, or be
propagating through a main-sequence wind which is also axisymmetric.
The other possibility is that the SNR's complex appearance results from
the inhomogeneous ISM into which it is expanding. In this context, an
unusual linear feature running north-south through the remnant may
represent a low-density tunnel which has been re-energised by encounter
with the SN shock, while a partial ring of emission to the north-east
may indicate a break-out into an adjacent bubble. A detailed
\HI\ emission study of the region, using a combination of compact ATCA
configurations and single-dish data, may give further insight into the
remnant and its environment.

We also detect the pulsar PSR~J1157--6224, 13 arcmin from the remnant's
centre, and find its measured properties to agree with those found in
previous observations.  On the basis of its \HI\ absorption, rotation
measure and dispersion measure, we argue that the pulsar is at a
distance of $\sim$10~kpc, compatible with the distance to the SNR.
However, the pulsar's large timing age and large displacement from the
remnant's centre make a physical association highly unlikely.  

The pulsar is 12 per cent circularly polarized and its detection through such
emission is straightforward.  Other sources in the field show no
circular polarization and the nature of the ATCA means that
instrumental effects in Stokes V are minimal.  This suggests that detection
through Stokes V is a useful method for finding new pulsars.

\section*{Acknowledgments}

We thank Jim Caswell, John Dickey, Vince McIntyre and Brad Wallace for
helpful discussions, and B\"{a}rbel Koribalski for supplying \HI\ data
on PSR~J1157--6224. Bob Sault, Richard Gooch and Neil Killeen assisted
with data reduction.  BMG acknowledges the support of an
Australian Postgraduate Award.  The Australia Telescope is funded by
the Commonwealth of Australia for operation as a National Facility
managed by CSIRO. This research has made use of the NASA Astrophysics
Data System and the CDS SIMBAD database.

\bibliographystyle{mn}
\bibliography{modrefs,psrrefs}
\label{lastpage}

\end{document}